\documentclass{caosp309}

\usepackage{graphicx}

\usepackage{natbib}
\articleNo{B20}
\pubyear{2024}
\volume{54}
\volnumber{1}
\firstpage{1}
\received{October 31, 2023}
\accepted{???}

\begin{document}

\htitle{Photometric monitoring of PMS stars}
%\hauthor{E.\,Semkov et al.}
\hauthor{E.\,Semkov, S.\,Ibryamov, S.\,Peneva and A.\,Mutafov}

\title{Photometric monitoring of PMS stars with the telescopes at Rozhen Observatory}

\author{
	E.\,Semkov\inst{1}\orcid{0000-0002-1839-3936}
      \and	
        S.\,Ibryamov\inst{2}\orcid{0000-0002-4618-1201}
      \and 
        S.\,Peneva\inst{1}   
       \and 
        A.\,Mutafov\inst{1} 
			}

\institute{
	   Institute of Astronomy and National Astronomical Observatory, Bulgarian Academy of Sciences, 72, Tsarigradsko Shose Blvd., 1784 Sofia, Bulgaria, \email{esemkov@astro.bas.bg}\\
         \and
           Department of Physics and Astronomy, Faculty of Natural Sciences, University of Shumen, 115, Universitetska Str., 9712 Shumen, Bulgaria \\
          }
\date{October 31, 2023}

\maketitle

\begin{abstract}
For several decades we have been performing photometric monitoring of some of the star formation regions. 
Significant place in our program take observations of objects of the type FU Orionis, EX Lupi, UX Orionis and other similar but unclassified objects. 
These three types of young variable objects show changes in brightness with large amplitudes and attract the attention of star formation researchers. 
But it is not always possible to distinguish them from each other without the presence of long-term multicolor photometric data. 
For this reason, we collect data from current CCD observations and supplement them with data from the photographic plates archives.
In this paper, we show the latest data from optical photometric studies of four PMS objects (V2493 Cyg, V582 Aur, V733 Cep and V1180 Tau) made at the Rozhen Observatory. 
Our monitoring is carried out in $BVRI$ filters, which allows studying the variability in color indexes also. 
By analysis the historical light curves of these objects we are trying to obtain information about the processes associated with the early stages of stellar  evolution.                                                                                            
\end{abstract}

\section{Introduction}

The photometric variability of pre-main sequence (PMS) stars is their main characteristic.
The processes that took place during the formation of stars are manifested by changes in brightness with large amplitudes, which are accessible to observations with telescopes of medium and small size.
Thus, for example, the recorded outburst of stars from the type of FU Orionis (FUor) reach amplitudes of up to 6 magnitudes (Audard et al. 2014).
These brightness changes are caused by an increase in the rate of accretion from the circumstellar disk onto the stellar surface (Hartmann \& Kenyon 1996).
Also the eclipses caused by clouds of dust, which are characteristic of stars of the UX Orionis (UXor) type, have amplitudes of the order of 3-4 stellar magnitudes (Nata et al. 1997).
On the other hand, the study of the processes taking place during star formation gives us valuable information about the evolution of stars and the accumulation of their mass.

\section{Observations}
National Astronomical Observatory Rozhen (Bulgaria) is situated in the Rhodope Mountains at 1750 m altitude and it is a leading astronomical center in the South-East Europe. 
Since its inception, the observatory has been equipped with three telescopes for optical observation: 2-m RCC telescope, 50/70 cm Schmidt telescope and 60-cm Cassegrain telescope. 
From mid-2023, the observatory already has a new 1.5-m RC robotic telescope, which will also be used for photometric observations.
For our photometric monitoring of PMS stars, we have mainly used the 2-m RCC and the Schmidt telescope of Rozhen observatory.
Additional information on the procedure for obtaining and analyzing the photometric data, as well as on the technical parameters of the CCD cameras used, can be found in Mutafov et al. (2022).
In order to construct the historical light curves of the studied objects, we have also used data from archival photographic observations where possible.

\section{Results and Discussion}

In this section we will present some of the most interesting results obtained from observations of PMS objects.
We have selected four stars that exhibit large-amplitude variability and are readily accessible for observations with small and medium-sized telescopes.

\subsection{V2493 Cyg}

An outburst of a FUor-type star located in the region of the North America Nebula (NGC 7000) was registered at the Rozhen Observatory in 2010 (Semkov et al. 2010).
The eruptive star received designation V2493 Cyg, but it has been known in previous studies as HBC 722 (Herbig et al. 1988). 
The star's brightness rises by over 4 mag. (R) within a few months, followed by an extended period of position at maximum light. (Semkov et al. 2021).
Spectral observations of V2493 Cyg show significant changes in the profiles of the Balmer hydrogen lines and the sodium doublet, which are characteristic of FUor objects (Miller et al. (2011); Semkov et al. (2012)).

Data from our monitoring of V2493 Cyg show photometric behavior that is not typical of the other FUor objects.
Usually, after the rapid increase in brightness, the FUor objects undergo a period of slow, gradual decrease in brightness over decades.
But in the case of V2493 Cyg, we observe a prolonged photometric plateau, with no indication of a significant decline in brightness.
Photometric data we collected from archival photographic observations do not indicate the presence of other outbursts, but only small-amplitude variability that is characteristic of T Tauri stars.

\begin{figure}
\centerline{\includegraphics[width=0.85\textwidth,clip=]{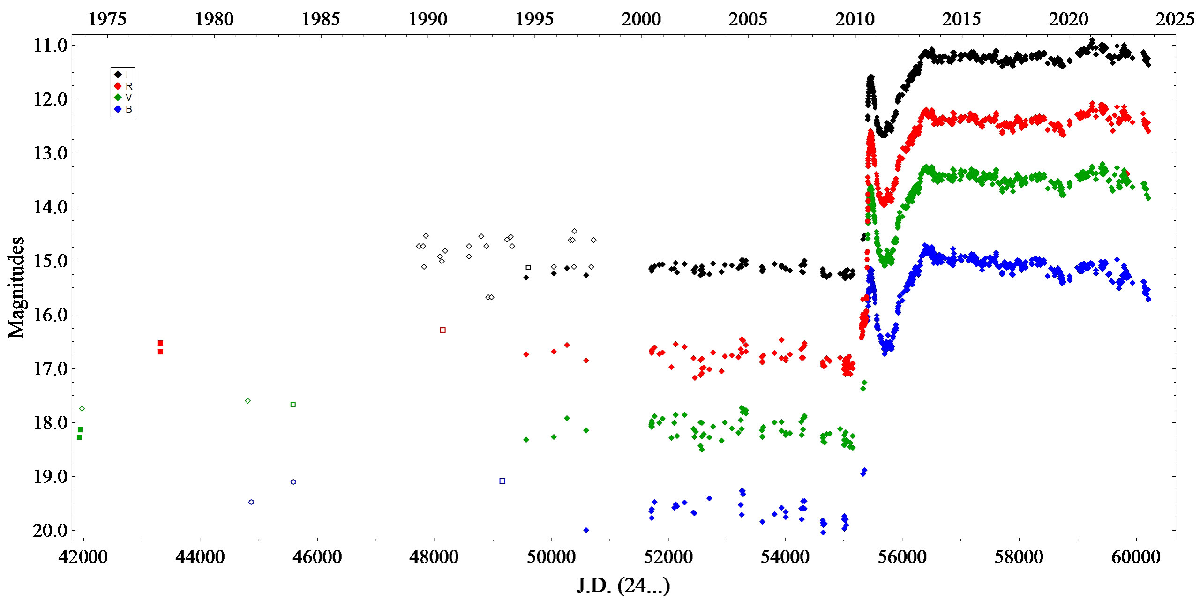}}
\caption{BVRI light curves of V2493 Cyg for the period 1973 September $-$ 2023 September. The symbols used are as in Semkov et al. (2012).}
\label{V2493Cyg}
\end{figure}

\subsection{V582 Aur}

The PMS star V582 Aur is located in a region of active star formation near Auriga OB2 association.
The star was registered in 2009 by amateur astronomer Anton Khruslov, as an object with multiple increased brightness compared to the maps of the Palomar Sky Survey.
Our data collected from photographic observations indicate that the outburst of V582 Aur began around 1986 (Semkov et al. 2013).
The results of our detailed photometric and spectral study of V582 Aur indicate that it has all observational characteristics of FUor objects.

\begin{figure}
\centerline{\includegraphics[width=0.85\textwidth,clip=]{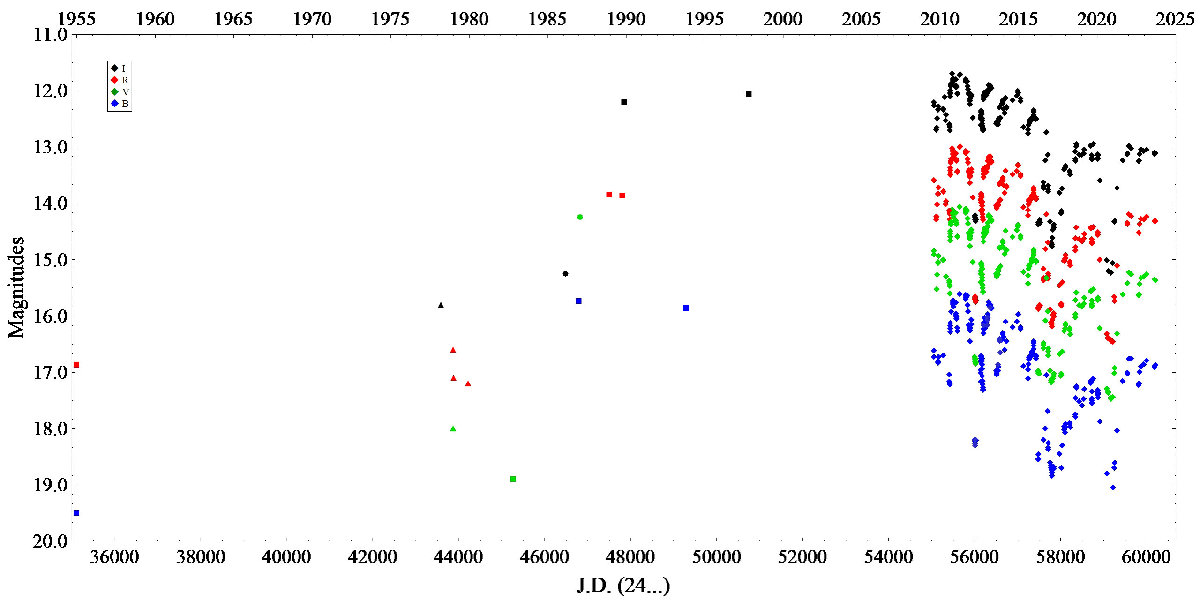}}
\caption{Historical BVRI light curves of V582 Aur for the period 1954 December $-$ 2023 September. The symbols used are as in Semkov et al. (2013).} 
\label{V582Aur}
\end{figure}
 
The results collected from our photometric monitoring of V582 Aur (2009-2023) show extremely strong photometric variability.
Along with this, significant changes in the spectrum of the star are also observed.
When the brightness of the star is in the high levels the spectrum is similar to a FUor object.
And accordingly, at low brightness levels, the spectrum of the V582 Aur is similar to the spectrum of a T Tauri star.
The explanation for this spectral and photometric variability is a combination of variable accretion from the circumstellar disc and variable extinction from the circumstellar environment (Semkov et al. (2013); Zsidi et al. (2019)).

\subsection{V733 Cep}

The variability of V733 Cep was discovered by Swedish amateur astronomer Roger Persson in 2004 after comparing the plate scans from the first and the second Palomar Sky
Survey. 
The first detailed study of V733 Cep is presented by Reipurth et al. (2007) who suggested that the outburst occurred in the period 1953-1984.
On the basis of spectral and photometric observations in the infrared region, Reipurth et al. (2007) show that this PMS star has all the main characteristics of FUor-type objects.

\begin{figure}
\centerline{\includegraphics[width=0.85\textwidth,clip=]{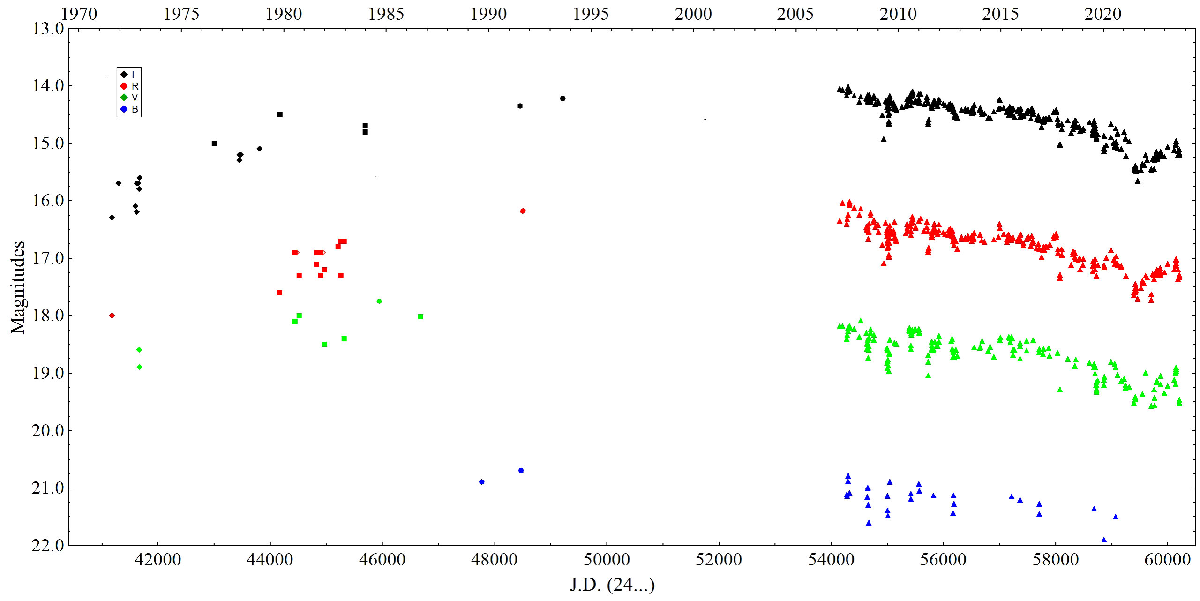}}
\caption{Historical BVRI light curves of V733 Cep for the period 1971 August $-$ 2023 September. The symbols used are as in Peneva et al. (2010).}
\label{V733Cep}
\end{figure}

Our paper (Peneva et al. 2010) containing data on the star from photometric monitoring and archival observations also showed a large-amplitude outburst.   
The rise in brightness has continued over the period 1971-1993 after which the V733 Cep reached its maximum brightness.
The amplitude registered by us in relation to the level of the Red map of the Palomar Sky Survey (1953) exceeds 4$\fm$5 (R). 
Subsequent photometric observations of V733 Cep show a gradual decrease in brightness, and over a period of 16 years (2007-2023) it has decreased by about 1$\fm$5 (R).    
Thus, the star is a unique FUor object in which a nearly symmetric light curve is observed, the rate of increase in brightness being similar to the rate of decrease in brightness.

\subsection{V1184 Tau}

A variable PMS star of unclear classification was discovered by Yun et al. (1997) in the field of the Bok globule CB 34.
Comparing the photometric observations obtained by the authors in 1991 with the plates from Palomar Sky Survey (1951) shows a change in the brightness of this star of about 3$\fm$7 (R).
The first guess of Yun et al. (1997) is that the star is a FUor object, but they don't rule out the possibility that they have discovered a new eclipsing variable star.

\begin{figure}
\centerline{\includegraphics[width=0.85\textwidth,clip=]{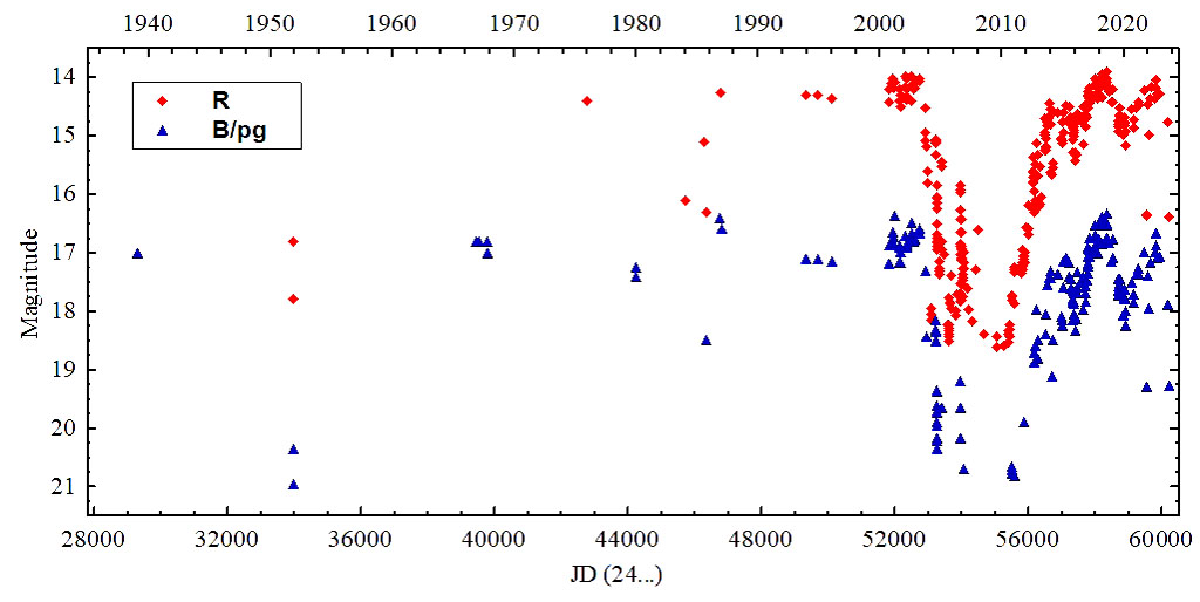}}
\caption{Historical R and B/pg light curves of V1184 Tau.}
\label{fsinus}
\end{figure}

We began regular photometric observations of this object (designated V1184 Tau) in 2000, with the first results showing variability characteristic of Classical T Tauri stars (Semkov 2003).
But in 2003, a deep eclipse of the star's brightness began, which lasted fourteen years (Semkov et al. 2015).
The amplitude of the eclipse reached $\Delta$$I$$\approx$$4\fm8$, and the changes in the color indices showed a color reversal (co-called "blueing"), characteristic of UX Orionis stars at low brightness levels.
This color-reversal effect is an evidence that the dimming was caused by dust clouds covering the star along the line of sight.
In recent years, V1184 Tau has undergone several more short eclipses, which indicate a strong inhomogeneity of the material covering the star.

%%%%%%%%%%%%%%%%%%%%%%%%%%%%%%%%%%%%%%%%%%%%%%%%%%%%%%%%%%%%%%%%%%%%%%%%%%%%%
%             B i b T e X   I N   C I T A T I O N S
% CAOSP BibTeX style (caosp.bst) is a modified style of A&A and APJ.
% That is why the rules regarding BibTeX can be found at:
%    http://ftp.edpsciences.org/pub/aa/aadoc.pdf (chapter 3.6)
%    http://ftp.edpsciences.org/pub/aa/bibtex/natnotes.pdf
% If you want to add in-text citation clickers that link to the
% corresponding ADS abstract pages, you can use the "TIPS" given at
%    http://ftp.edpsciences.org/pub/aa/readme.html
% Next lines show the examples of basic BibTeX usage in citations:
% The \cite command functions as follows:
%   \citet{key} ==>>                Jones et al. (1990)
%   \citet*{key} ==>>               Jones, Baker, and Smith (1990)
%   \citep{key} ==>>                (Jones et al., 1990)
%   \citep*{key} ==>>               (Jones, Baker, and Smith, 1990)
%   \citep[chap. 2]{key} ==>>       (Jones et al., 1990, chap. 2)  
%   \citep[e.g.][]{key} ==>>        (e.g. Jones et al., 1990)
%   \citep[e.g.][p. 32]{key} ==>>   (e.g. Jones et al., p. 32)
%   \citeauthor{key} ==>>           Jones et al.
%   \citeauthor*{key} ==>>          Jones, Baker, and Smith
%   \citeyear{key} ==>>             1990
%
% For an advanced BibTeX usage (muliple citations, etc.) an author is
% advised to consult the reference sheet of the natbib package at:   
%  http://ftp.edpsciences.org/pub/aa/bibtex/natnotes.pdf
%
%%%%%%%%%%%%%%%%%%%%%%%%%%%%%%%%%%%%%%%%%%%%%%%%%%%%%%%%%%%%%%%%%%%%%%%%%%%%%

\acknowledgements
This research has made use of NASA’s Astrophysics Data System Abstract Service. 
The research infrastructure this research was done with is  funded by the Ministry of Education
and Science of Bulgaria (support for the Bulgarian National Roadmap for Research Infrastructure).

\end{document}